\documentclass[a4paper,12pt,english]{article}
\usepackage{amsfonts}
\usepackage{graphicx}
\usepackage{amsmath}
\usepackage{color}

\newcommand{\be}{\begin{equation}}
\newcommand{\ee}{\end{equation}}
\input{tcilatex}

\begin{document}


\begin{titlepage}
\begin{center}

\noindent{{\LARGE{The large $D$ limit of dimensionally continued gravity}}}

\smallskip
\smallskip

\smallskip
\smallskip
\smallskip
\smallskip
\noindent{\large{Gaston Giribet}}

\smallskip
\smallskip

\end{center}
\smallskip
\smallskip
\centerline{Physics Department, University of Buenos Aires, and IFIBA-CONICET}
\centerline{{\it Ciudad Universitaria, Pabell\'on 1, 1428, Buenos Aires, Argentina.}}
\smallskip
\smallskip
\smallskip
\smallskip

\bigskip

\bigskip

\bigskip

\bigskip

\begin{abstract}
In a recent paper \cite{Emparan} Emparan, Suzuki, and Tanabe studied general relativity in the limit in which the number of spacetime dimensions $D$ tends to infinity. They showed
that, in such limit, the theory simplifies notably. It reduces to a theory whose fundamental objects, black holes and black branes, behave as
non-interacting particles. Here, we consider a different way of extending gravity to $D$ dimensions. We present a special limit of dimensionally continued gravity
in which black holes retain their gravitational interaction at large $D$ and still have entropy proportional to the mass. The similarities and differences with the limit considered in \cite{Emparan} are discussed.

\end{abstract}

\end{titlepage}


\section{Introduction}

There exist several ways of extending general relativity (GR) to higher
dimensions. The simplest one is retaining the form of Einstein-Hilbert
Lagrangian density and then extend the action to $D\geq 4$ dimensions.
However, this proposal encounters a naturalness problem since in $D>4$
dimensions Einstein tensor is not as special as it is in $D=4$. In $D>4$,
the requirement of the equations of motion to be symmetric rank-two
covariantly conserved equations of second order does not select Einstein
tensor uniquely. In addition, there exists the possibility to supplement
Einstein-Hilbert action with dimensionally extended characters of the form%
\footnote{%
Here we will work in the first order formalism; see Section 2 for
conventions.}%
\begin{equation}
\chi _{n}=\int \varepsilon
_{a_{1}a_{2}...a_{2n}...a_{D}}R^{a_{1}a_{2}}\wedge R^{a_{3}a_{4}}\wedge
...R^{a_{2n-1}a_{2n}}\wedge e^{a_{2n+1}}\wedge e^{a_{2n+2}}\wedge
...e^{a_{D}},  \label{X}
\end{equation}%
which, despite of being of order $R^{n}$, yield second-order field
equations. Then, it is natural to inquire about why not to include the whole
hierarchy of characters $\chi _{n}$ up to order $(D-1)/2$ in the gravity
action. Similarly to how Einstein-Hilbert action $\chi _{1}$ can be thought
of as the dimensional extension of Euler characteristic in $D=2$ dimensions,
in $D>4$ it is natural to define the gravity action by including the
dimensional extension of the other Chern-Weil topological invariants. In $%
D=4 $, for instance, the Gauss-Bonnet theorem implies that $R^{2}$ terms of
this sort do not modify Einstein equations, as early noticed by Lanczos \cite%
{Lanczos}; however, in $D>4$ it is natural to include such terms. The same
happens with $\chi _{n}$ in higher dimensions. The theory of gravity in $D$
dimensions whose action consists of all the dimensionally extended
topological densities (\ref{X}) up to $n=(D-1)/2$ is known under the rubric
of Lovelock, after D. Lovelock have found in \cite{Lovelock} the
generalization of the Einstein tensor to $D$ dimensions.

This digression about which is the natural extension of GR to $D$ dimensions
acquires particular importance in relation to recent studies on the behavior
of gravity in the large $D$ limit \cite{Emparan}. This limit had already
been considered in the literature, for instance in Refs. \cite{Strominger,
Canfora, otro, otro2}, and it was recently revisited in \cite{Emparan} by
Emparan et al., who observed that GR simplifies notably when $D$ goes to
infinity. In particular, they observed that in this limit the theory reduces
to a theory of non-interacting particles. The fundamental objects of the
theory, black holes and black $p$-branes, exhibit vanishing cross-section
and behave like dust matter.

The idea of considering the large $D$ limit of gravity theory can be
motivated by the large $N$ limit of gauge theories. The latter has shown to
be a fruitful tool to investigate the structure of both Yang-Mills and
Chern-Simons theories. Exceptis excipiendis, gravity theory can also be
considered as a gauge theory for the local Lorentz group $SO(D-1,1)$. In
turn, it is natural to explore whether one can extract relevant information
from studying its $1/D$ expansion. Of course, besides the mathematical
analogy with the large $N$ limit of gauge theories, the fact that $D$
represents the dimensionality of the spacetime itself introduces additional
conceptual difficulties. Nevertheless, as explained in \cite{Emparan}, this
limit may still be considered and interesting physical information can be
extracted from studying it.

Here, we will consider a different way of extending gravity to $D$
dimensions and study the limit of large $D$. More precisely, we will
consider the gravity theory defined by the action that includes all terms (%
\ref{X}) up to a given order $R^{k}$, with $k\leq (D-1)/2$. For this type of
theories, the mentioned analogy between the large $N$ limit of gauge
theories and the large $D$ limit of gravity is even more direct since in the
particular case $2k+1=D$ the actions we will consider coincide with
Chern-Simons actions (CS) for the gauge group $SO(D-1,2)$, and then they
correspond to actual gauge theories. This can be regarded as an additional
motivation to study these models. For $3<2k+1<D$, instead, one is in an
intermediate situation, between GR\ and CS. This will allow us to play
between two extremes, between $k=1$ and $k=(D-1)/2$. The fact of having now
two parameters, $D$ and $k$, allows us to take the large $D$ limit in
different manners. For instance, we can take $D$ going to infinity by
keeping $k$ fixed, but we also can take both $D$ and $k$ large in such a way
that the quotient $D/k$ remains fixed. In the latter case we will find that,
contrary to the limit considered in \cite{Emparan}, the black holes happen
to retain their gravitational potential in a finite region outside the
horizon. At first, this might sound surprising since the $R^{k}$ terms of
Lovelock theory are expected to introduce ultraviolet effects merely. In the
words of \cite{Emparan}, the fact that Riemann curvature tends to strongly
localize close to the horizon indicates that the dust picture should still
apply [in Lovelock theory] at least in some situations. We will see that,
although this is the case in certain situations, it is not true in general
and Lovelock black holes may actually retain the interactions at large $D$.

\section{Dimensionally continued gravity}

As said, we will be concerned with Lovelock theory of gravity. The idea of
considering Lovelock theory in relation to the large $D$ limit of gravity
was already proposed in \cite{Emparan}. The action of the theory can be
written as follows%
\begin{equation}
S=\kappa ^{-1}\dsum\nolimits_{n=0}^{D/2}\alpha _{n}\chi _{n}  \label{S}
\end{equation}%
where the terms $\chi _{n}$ are given by%
\begin{equation}
\chi _{n}=\int \varepsilon
_{a_{1}a_{2}...a_{2n}...a_{D}}R^{a_{1}a_{2}}\wedge R^{a_{3}a_{4}}\wedge
...R^{a_{2n-1}a_{2n}}\wedge e^{a_{2n+1}}\wedge e^{a_{2n+2}}\wedge
...e^{a_{D}}  \label{L}
\end{equation}%
where $R^{ab}=R_{\ \mu \nu }^{ab}dx^{\mu }\wedge dx^{\nu }$ is the curvature
two-form, $R^{ab}=d\omega ^{ab}+\omega _{\ c}^{a}\wedge \omega ^{cb}$, with $%
\omega ^{ab}=\omega _{\mu }^{ab}dx^{\mu }$ being the spin connection
one-form, and $e^{a}=e_{\mu }^{a}dx^{\mu }$ is the vierbein one-form. Latin
indices refer to indices in the tangent bundle while Greek indices refer to
indices in the spacetime. In (\ref{S}) $\kappa $ and $\alpha _{n}$ are
dimensionful constant that introduce new fundamental scales in the theory.
We will discuss these scales below.

The equations of motion are obtained by varying (\ref{S}) with respect to
the vierbein and the spin connection. Varying with respect to $e^{a}$ yields 
\begin{equation}
\dsum\nolimits_{n=0}^{D/2}\alpha _{n}(D-2n)\varepsilon
_{a_{{}}a_{2}a_{3}...a_{D}}R^{a_{2}a_{3}}\wedge ...R^{a_{2n}a_{2n+1}}\wedge
e^{a_{2n+2}}\wedge ...e^{a_{D}}=0,  \label{UNO}
\end{equation}%
while varying with respect to $\omega ^{ab}$ yields%
\begin{equation}
\dsum\nolimits_{n=0}^{D/2}\alpha _{n}n(D-2n)\varepsilon
_{aba_{3}a_{4}...a_{D}}R^{a_{3}a_{4}}\wedge ...R^{a_{2n-1}a_{2n}}\wedge
T^{a_{2n+1}}\wedge e^{a_{2n+1}}\wedge ...e^{a_{D}}=0,  \label{DOS}
\end{equation}%
where $T^{a}=de^{a}+\omega _{\ b}^{a}\wedge e^{b}$ is the torsion two-form.
Equations (\ref{DOS}) vanish if torsion is taken to be zero. Notice this is
sufficient but not necessary condition if $D\geq 4$. Here we will consider $%
T^{a}=0$. Then, the equations that remain to be solved are (\ref{UNO}).

In addition to considering (\ref{UNO}) we will define our theory by
specifying a criterion to choose special sets of coupling constants $\alpha
_{n}$. We will follow the criterion of Ref. \cite{scan}. That is, we will
demand the theory to admit a unique maximally symmetric vacuum. This
prevents the theory from suffering from ghost instabilities \cite%
{BoulwareDeser} and other type of pathologies \cite{Gomberoff}. This
requirement of a unique vacuum leads to the following choice of couplings
constants \cite{scan}%
\begin{equation}
\alpha _{n\leq k}=\frac{L^{2(n-k)}}{(D-2n)}\frac{\Gamma (k+1)}{\Gamma
(n+1)\Gamma (k-n+1)},\qquad  \label{alpha}
\end{equation}%
while $\alpha _{n>k}=0$. Ipso facto, this introduces an additional parameter
of the theory, $k$, which represents the highest order $R^{k}$ in the
action. This invites to define the critical dimension $D_{c}\equiv 2k+1$,
which represents the minimum number of dimensions such that a term $\chi
_{k} $ in the action would contribute non-trivially to the equations of
motion. In other words, $\chi _{k}$ is the Chern-Weil topological invariant
in $D_{c}-1$ dimensions. In the particular case $D=D_{c}$ (i.e. $D=2k+1$)
the theory defined by (\ref{S})-(\ref{alpha}) coincides with the
Chern-Simons theory of gravity \cite{Zanelli}. In the case $D=D_{c}+1$ the
action admits to be written as a Pfaffian, and then it is often referred to
as the Born-Infeld action \cite{BTZ2}. Hereafter, we will be viewing the
gravity theory as a biparametric model, and consequently we will express all
the formulae below as functions of $D$ and $D_{c}$.

At first glance it might seem remarkable that demanding the theory to admit
a unique maximally symmetric vacuum yields a relation between the coupling
constants $\alpha _{n}$ that makes all of them to be determined by a unique
fundamental scale $L$. However, due to the plethora of vacua in
higher-curvature theory, such a requirement turns out to be actually very
restrictive and this is why, apart from Planck scale $\kappa $, $L$ appears
as the only relevant scale.

About Planck scale, we find convenient to define Newton constant as follows%
\begin{equation}
\kappa =2\Gamma (D-1)\Omega _{D-2}G_{D,D_{c}}  \label{torba}
\end{equation}%
where $G_{D,D_{c}}$ has dimensions of (length)$^{D-D_{c}+1}$, such that the
coefficient of the Einstein-Hilbert term, $\alpha _{1}/\kappa $, has
dimensions of (length)$^{2-D}$ as required. In (\ref{torba}), 
\begin{equation}
\Omega _{D-2}=\frac{2\pi ^{(D-1)/2}}{\Gamma (\frac{D-1}{2})}
\end{equation}%
is the volume of the unit ($D-2$)-sphere.

We also recognize the cosmological constant%
\begin{equation}
\Lambda =-\frac{(D-1)(D-2)}{2L^{2}},
\end{equation}%
which is given by the coefficient $\alpha _{0}/\kappa $ in the action above.

\section{Dimensionally continued black holes}

\subsection{Classical black holes}

Another interesting features of the set of theories defined by the choice (%
\ref{alpha}) is the fact that they can be solved analytically in a variety
of examples. In particular, their spherically symmetric solutions can be
found explicitly for generic values of $D$ and $D_{c}.$ These metrics take
the form \cite{scan}%
\begin{equation}
ds^{2}=-fdt^{2}+f^{-1}dr^{2}+r^{2}d\Omega _{D-2}^{2}  \label{ds}
\end{equation}%
with 
\begin{equation}
f(r)=1+\frac{r^{2}}{L^{2}}-\left( \frac{r_{0}}{r}\right)
^{2(D-D_{c})/(D_{c}-1)}.  \label{f}
\end{equation}

In the particular case $D_{c}=3$ ($k=1$) this solution reduces to
Schwarzschild-Tangherlini solution of GR, as expected. In the cases $D=D_{c}$%
, on the other hand, this solution coincides with the Ba\~{n}%
ados-Teitelboim-Zanelli solution for Chern-Simons gravity \cite{BTZ}.

The mass of solutions (\ref{ds})-(\ref{f}) can be computed by resorting to
the Hamiltonian formalism \cite{scan}. The result is expressed in terms of
the horizon radius $r_{H}$ as follows%
\begin{equation}
M=\frac{r_{H}^{D-D_{c}}}{2G_{D,D_{c}}}\left( 1+\frac{r_{H}^{2}}{L^{2}}%
\right) ^{(D_{c}-1)/2}  \label{M}
\end{equation}%
up to an additive constant that can be set to zero for simplicity.

At this stage we are ready to study the geometry of these black holes in the
large $D$ limit. In this limit the volume of the ($D-2$)-sphere exhibits the
Stirling scaling $\Omega _{D-2}\sim D^{-D/2}$, so that it tends to zero.
This means that the base manifold of the black hole shrinks in the large $D$
limit. This was rephrased in \cite{Emparan} as the black holes having
vanishing cross-section when $D$ goes to infinity.

Outside the horizon, the gravitational potential damps off faster as $D$
increases. This implies that the gravitational interactions between
Schwarzschild-Tangherlini black holes extinguishes in the large $D$ limit.
In the general case (\ref{f}), the way the gravitational potential scales
with $D$ also depends on how $D_{c}$ scales. If $D_{c}$ remains finite in
the large $D$ limit, the behavior of solutions (\ref{ds})-(\ref{f}) would be
qualitatively similar to that of \cite{Emparan}. However, if, instead, both $%
D$ and $D_{c}$ are taken to infinity in a way that the quotient $D/D_{c}$
remains fixed, then the black holes happen to retain their gravitational
interaction outside the horizon. In this limit $\Omega _{D-2}$ still
vanishes, but metric function (\ref{f}) has a large $D$ behavior%
\begin{equation}
f(r)\simeq 1+\frac{r^{2}}{L^{2}}-\left( \frac{r_{0}}{r}\right)
^{2(D/D_{c}-1)},
\end{equation}%
and the gravitational potential remains finite.

\subsection{Quantum black holes}

Now, let us turn to discuss black holes in the quantum regime. The Hawking
temperature associated to black holes (\ref{ds})-(\ref{f}) can easily be
calculated to be%
\begin{equation}
T=\frac{\hbar }{2\pi (D_{c}-1)}\left( \frac{(D-1)r_{H}}{L^{2}}+\frac{%
(D-D_{c})}{r_{H}}\right) ,  \label{T}
\end{equation}%
which reproduces the GR result for $D_{c}=3$. We observe that the theory for
generic $D$ and $D_{c}$ seems to exhibit Hawking-Page transition, provided $%
L $ is finite. If $D$ goes to infinity and $D_{c}$ remains fixed,
temperature (\ref{T}) diverges. Still, there is a point at which the
specific heat changes its sign and the transition occurs. This happens at
the scale $r=L\sqrt{(D-D_{c})/(D-1)}\simeq L.$ On the other hand, in
contrast to what happens in GR, the presence of higher-curvature terms
permits to take the large $D$ limit in a way that $T$ remains finite. This
is achieved by taking $D_{c}$ to infinity as well by keeping $D/D_{c}$
fixed. For instance, if we define $D_{c}=D(1-\alpha )$, then the scale at
which the transition takes place is governed by $\alpha $, obtaining $%
r\simeq L\sqrt{\alpha }$.

Let us study the case of asymptotically flat solutions. This is obtained by
taking the large $L$ limit. In the theories defined by (\ref{S})-(\ref{alpha}%
) this corresponds to having only the highest curvature term $R^{k}$ turned
on. In this limit, we find%
\begin{equation}
T=\frac{\hbar (D-D_{c})}{2\pi (D_{c}-1)r_{H}}.  \label{T2}
\end{equation}

The entropy, on the other hand, is%
\begin{equation}
S=\frac{\pi r_{H}^{D-D_{c}+1}(D_{c}-1)}{\hbar G_{D,D_{c}}(D-D_{c}+1)}.
\label{S2}
\end{equation}

Because of the presence of higher-curvature terms in the action, these black
holes happen not to obey the Bekenstein-Hawking area law. Instead, entropy
is a different monotonic function of the horizon area $A$, namely $S\propto
A^{\frac{D-D_{c}+1}{D-2}}$. From (\ref{S2}) and (\ref{M}) we also observe
that even in the particular limit in which the black holes retain their
gravitational potential, the entropy and the mass go $S\propto M$ when $D$
is large. This implies that such a behavior is not necessarily associated to
the non-interacting picture, at least not in a simple way.

\subsection{Black $p$-branes}

The study of the thermodynamics of black holes (\ref{ds})-(\ref{f}) enables
to study the thermodynamical stability of other black objects of the theory.
For instance, consider black $p$-branes. That is, consider solutions of the
form $\Sigma _{D-p}\times T^{p}$, with $T^{p}$ being a $p$-torus and $\Sigma
_{D-p}$ being a black hole of the type discussed above. This type of
solutions was considered in Refs. \cite{branes, KastorMann}, where it was
shown that metric%
\begin{equation*}
ds^{2}=-fdt^{2}+f^{-1}dr^{2}+r^{2}d\Omega
_{D-2-p}^{2}+\dsum\nolimits_{i=1}^{p}dz_{i}^{2}
\end{equation*}%
with%
\begin{equation*}
f(r)=1-\left( \frac{r_{0}}{r}\right) ^{2(D-p-D_{c})/(D_{c}-1)}
\end{equation*}%
are solutions of the theory (\ref{S})-(\ref{alpha}) in the limit $%
L\rightarrow \infty $.

One can analyze the thermodynamical instability of black $p$-branes by
comparing the entropy of such a configuration with that of a black hole.
This requires a careful analysis of the parameters involved in each
configuration when comparing them in the microcanonical ensemble. The
thermodynamical stability analysis yields the following result for the
quotient of entropies \cite{branes}%
\begin{equation}
\frac{S^{\text{Black }p\text{-brane}}}{S^{\text{Black hole}}}=\frac{%
(D-D_{c}+1)}{(D-D_{c}-p)}(2G_{D,D_{c}})^{\lambda _{1}}M^{\lambda
_{2}}(A_{D,D_{c},p})^{\lambda _{3}},  \label{j1}
\end{equation}%
with%
\begin{equation}
A_{D,D_{c},p}=\frac{\Gamma (D-D_{c}-p+1)\Gamma (D-1)\Gamma ((D-p-1)/2)}{%
\Gamma (D-D_{c}+1)\Gamma (D-1-p)\Gamma ((D-1)/2)}\frac{\pi ^{p/2}}{Vol(T^{p})%
}  \label{j2}
\end{equation}%
and with critical exponents%
\begin{eqnarray}
\lambda _{1} &=&\frac{1}{D-D_{c}-p}-\frac{1}{D-D_{c}}  \label{a1} \\
\lambda _{2} &=&\frac{D-D_{c}+1-p}{D-D_{c}-p}-\frac{D-D_{c}+1}{D-D_{c}}
\label{a2} \\
\lambda _{3} &=&\frac{1}{D-D_{c}-p}  \label{a3}
\end{eqnarray}

From (\ref{j1}) we observe that the thermodynamical analysis of the black
hole / black brane transition in this theory is qualitatively similar to
that of GR: There always exists a critical mass above which the black $p$%
-brane is the preferable configuration. The natural question arises as to
how this picture is modified in the large $D$ limit. For instance, in the
large $D$ limit with $D_{c}\ $fixed, all the exponents $\lambda _{1,2,3}$
tend to zero. This behavior is actually expected because here we are
considering $p$ fixed. A similar behavior is exhibited also in the limit in
which the quotient $D/D_{c}$ remains fixed. An interesting limit is given by
taking both parameters to infinity by keeping the difference $D-D_{c}$
finite. In this limit, exponents $\lambda _{1,2,3}$ remain finite while $%
A_{D,D_{c},p}$ scales as $\sim D^{p/2}/Vol(T^{p})$. It would be interesting
to study the instability of $p$-brane solutions of this theory in a similar
way to what has been done in Refs. \cite{brana1, brana2} at large $D$. The
analysis of mechanical stability, on the other hand, can hardly be
accomplished for these theories. This is mainly because of two reasons:
First, the higher-curvature terms in the action introduce higher powers of
the derivatives that make the complexity of the equations to grow
dramatically even for large $D$. Secondly, the special theories that are
being selected by demanding (\ref{alpha}) have the property of having a
unique maximally symmetric vacuum, and this produces that the equations of
motion factorize in a way that the first orders in perturbation theory
identically vanish, making necessary to go beyond the linear approximation.
It would also be interesting to study the large $D$ limit of Lovelock theory
in relation to other issues as holographic applications, causality bounds 
\cite{Jose}, and other subjects in which this theory presents remarkable
curiosities as well. This is matter of further study.

\begin{equation*}
\end{equation*}%
This work was supported by ANPCyT, CONICET,\ and UBA. The author thanks
Julio Oliva and Ricardo Troncoso for previous collaboration in related
subjects, and thanks Andr\'{e}s Goya for discussions.

\end{document}